\documentclass[aps,twocolumn,superscriptaddress,showpacs]{revtex4-1}

\usepackage{graphicx}
\usepackage{dcolumn}
\usepackage{bm}
\usepackage{amsmath}
\usepackage{color}
\usepackage{ulem}
\usepackage{threeparttable}
\usepackage{enumerate}
\usepackage{paralist}

\begin{document}

\title{Absence of long-range order in an $XY$ pyrochlore antiferromagnet Er$_2$AlSbO$_7$}

\author{H. L. Che}
\affiliation{Department of Physics, Hefei National Laboratory for Physical Sciences at Microscale, and Key Laboratory of Strongly-Coupled Quantum Matter Physics (CAS), University of Science and Technology of China, Hefei, Anhui 230026, People's Republic of China}

\author{Z. Y. Zhao}
\email{zhiyingzhao@fjirsm.ac.cn}
\affiliation{State Key Laboratory of Structural Chemistry, Fujian Institute of Research on the Structure of Matter, Chinese Academy of Sciences, Fuzhou, Fujian 350002, People's Republic of China}

\author{X. Rao}
\affiliation{Department of Physics, Hefei National Laboratory for Physical Sciences at Microscale, and Key Laboratory of Strongly-Coupled Quantum Matter Physics (CAS), University of Science and Technology of China, Hefei, Anhui 230026, People's Republic of China}

\author{L. G. Chu}
\affiliation{Department of Physics, Hefei National Laboratory for Physical Sciences at Microscale, and Key Laboratory of Strongly-Coupled Quantum Matter Physics (CAS), University of Science and Technology of China, Hefei, Anhui 230026, People's Republic of China}

\author{N. Li}
\affiliation{Department of Physics, Hefei National Laboratory for Physical Sciences at Microscale, and Key Laboratory of Strongly-Coupled Quantum Matter Physics (CAS), University of Science and Technology of China, Hefei, Anhui 230026, People's Republic of China}

\author{W. J. Chu}
\affiliation{Department of Physics, Hefei National Laboratory for Physical Sciences at Microscale, and Key Laboratory of Strongly-Coupled Quantum Matter Physics (CAS), University of Science and Technology of China, Hefei, Anhui 230026, People's Republic of China}

\author{P. Gao}
\affiliation{Department of Physics, Hefei National Laboratory for Physical Sciences at Microscale, and Key Laboratory of Strongly-Coupled Quantum Matter Physics (CAS), University of Science and Technology of China, Hefei, Anhui 230026, People's Republic of China}

\author{X. Y. Yue}
\affiliation{Institute of Physical Science and Information Technology, Anhui University, Hefei, Anhui 230601, People's Republic of China}

\author{Y. Zhou}
\affiliation{Institute of Physical Science and Information Technology, Anhui University, Hefei, Anhui 230601, People's Republic of China}

\author{Q. J. Li}
\affiliation{School of Physics and Material Sciences, Institute of Physical Science and Information Technology, Anhui University, Hefei, Anhui 230601, People's Republic of China}

\author{Q. Huang}
\affiliation{Department of Physics and Astronomy, University of Tennessee, Knoxville, Tennessee 37996, USA}

\author{E. S. Choi}
\affiliation{National High Magnetic Field Laboratory, Florida State University, Tallahassee, Florida 32306-4005, USA}

\author{Y. Y. Han}
\affiliation{High Magnetic Field Laboratory, Chinese Academy of Sciences, Hefei, Anhui 230031, People¡¯s Republic of China}

\author{Z. Z. He}
\affiliation{State Key Laboratory of Structural Chemistry, Fujian Institute of Research on the Structure of Matter, Chinese Academy of Sciences, Fuzhou, Fujian 350002, People's Republic of China}

\author{H. D. Zhou}
\affiliation{Department of Physics and Astronomy, University of Tennessee, Knoxville, Tennessee 37996, USA}
\affiliation{National High Magnetic Field Laboratory, Florida State University, Tallahassee, Florida 32306-4005, USA}

\author{X. Zhao}
\affiliation{School of Physical Sciences, University of Science and Technology of China, Hefei, Anhui 230026, People's Republic of China}

\author{X. F. Sun}
\email{xfsun@ustc.edu.cn}
\affiliation{Department of Physics, Hefei National Laboratory for Physical Sciences at Microscale, and Key Laboratory of Strongly-Coupled Quantum Matter Physics (CAS), University of Science and Technology of China, Hefei, Anhui 230026, People's Republic of China}
\affiliation{Institute of Physical Science and Information Technology, Anhui University, Hefei, Anhui 230601, People's Republic of China}
\affiliation{Collaborative Innovation Center of Advanced Microstructures, Nanjing University, Nanjing, Jiangsu 210093, People's Republic of China}

\date{\today}

\begin{abstract}

Rare-earth pyrochlores are known to exhibit exotic magnetic phenomena. We report a study of crystal growth and characterizations of a new rare-earth compound Er$_2$AlSbO$_7$, in which Al$^{3+}$ and Sb$^{5+}$ ions share the same positions with a random distribution. The magnetism are studied by magnetic susceptibility, specific heat and thermal conductivity measurements at low temperatures down to several tens of milli-kelvin. Different from the other reported Er-based pyrochlores exhibiting distinct magnetically ordered states, a spin-freezing transition is detected in Er$_2$AlSbO$_7$ below 0.37 K, which is primarily ascribed to the inherent structural disorder. A cluster spin-glass state is proposed in view of the frequency dependence of the peak position in the ac susceptibility. In addition, the temperature and field dependence of thermal conductivity indicates rather strong spin fluctuations which is probably due to the phase competition.

\end{abstract}


\maketitle

\section{Introduction}

Geometrically frustrated antiferromagnets have been a topic of intensive interest due to their abundant and exotic magnetisms. The rare-earth pyrochlores are one of the famous families with three-dimensional spin frustration \cite{Review-1}. These materials have pyrochlore crystal structure with the space group $Fd\bar{3}m$, in which the magnetic rare-earth ions form a network of corner-sharing tetrahedra and are prone to a high degree of geometric frustration. However, these materials are greatly sensitive to weak perturbations (e.g. single-ion anisotropy, dipolar interaction or quantum fluctuations) beyond the nearest-neighboring exchange, which results in unconventional low-temperature magnetic and thermodynamic properties including classical spin ice, quantum spin ice, etc.

In this material family, Er-based pyrochlores with quantum $XY$ spin anisotropy have been found to exhibit some peculiar magnetic phenomena \cite{Review-XY pyrochlores-1, Review-XY pyrochlores-2, Review-XY pyrochlores-3, Review-XY pyrochlores-4}. The exchange interactions between the nearest-neighboring Er ions are anisotropic on account of the interplay between the spin-orbit coupling and crystal electric field (CEF), leading to the anisotropic exchange phase diagram \cite{Phase-diagram-1, Phase-diagram-2, Phase-diagram-3}. There are four types of $k =$ 0 magnetic structures, including a noncoplanar antiferromagnetic (AFM) state $\psi_2$ ($\Gamma_5$), a coplanar AFM state $\psi_3$ ($\Gamma_5$), a Palmer-Chalker (PC) state $\Gamma_7$, and a splayed ferromagnetic (FM) state $\Gamma_9$. Ground state of the Er-based pyrochlores is sensitive to the exchange parameter, and could be finely tuned by the chemical pressure. For example, Er$_2$Ti$_2$O$_7$ is a well-studied pyrochlore and a $\psi_2$ state is selected by the thermal or quantum fluctuations below $T\rm_N =$ 1.2 K \cite{Er2Ti2O7-1, Er2Ti2O7-2, Er2Ti2O7-3, Er2Ti2O7-4, Er2Ti2O7-5, Er2Ti2O7-6}, while Er$_2$Ge$_2$O$_7$ enters into the $\psi_3$ state below $T\rm_N =$ 1.4 K \cite{Er2Ge2O7-1,Er2Ge2O7-2}. Both Er$_2$Sn$_2$O$_7$ \cite{Er2Sn2O7-1, Er2Sn2O7-2} and Er$_2$Pt$_2$O$_7$ \cite{Er2Pt2O7-1, Er2Pt2O7-2}, however, undergo to the PC state below 0.1 K and 0.3 K, respectively. All these Er-based pyrochlores are magnetically ordered at low temperatures, but the drastic reduction of $T\rm_N$ for Er$_2$Sn$_2$O$_7$ and Er$_2$Pt$_2$O$_7$ is discussed to be originated from the phase competition between different ordered states in the anisotropic exchange phase diagram \cite{Er2Sn2O7-2,Er2Pt2O7-2}.

In this work, we report a new Er-based pyrochlore Er$_2$AlSbO$_7$ with inherent structural disorder, that is, Al$^{3+}$ and Sb$^{5+}$ ions share the same positions with a random distribution. With detailed characterizations by means of dc and ac susceptibility, magnetization, specific heat, and thermal conductivity measurements, the ground state of Er$_2$AlSbO$_7$ is found to be a spin-freezing state. The absence of magnetic order is dominantly ascribed to the exchange disorder on Er$^{3+}$ ions and a cluster spin-glass state is proposed. Moreover, the possible influence of phase competition on the suppression of magnetic order is also discussed.

\section{Experiments}

Er$_2$AlSbO$_7$ single crystals were grown by using the flux method. Mixture of Er$_2$O$_3$, Sb$_2$O$_3$, and Al$_2$O$_3$ powder with stoichiometric ratio were carefully ground. PbF$_2$ was used as flux and added into the mixture in a weight ratio of 12 : 1. The final mixture was loaded into an aluminum crucible, kept at 1200 $^\circ$C for 12 hours, and then slowly cooled to 800 $^\circ$C at a rate of 2 $^\circ$C/h before quickly cooled to room temperature at a rate of 100 $^\circ$C/h. The as-grown single crystals are pink and translucent with hexagon facet as shown in the inset to Fig. \ref{Structure}(a). The typical dimension of the crystal is about 1.5 $\times$ 1.5 $\times$ 1.5 mm$^3$. Lu$_2$AlSbO$_7$ single crystals used as a lattice reference in the specific heat analysis were grown in the similar way.

Single crystal x-ray diffraction (XRD) was performed on a Rigaku Mercury CCD diffractometer equipped with a graphite-monochromated Mo K$\alpha$ radiation ($\lambda$ = 0.71 {\AA}) at room temperature. DC magnetic susceptibility was measured by using a SQUID-VSM (Quantum Design). AC magnetic susceptibility was measured with the conventional mutual inductance technique with a oscillating field of 1 Oe at frequencies between 71 Hz and 1231 Hz at the temperature range from 20 mK to 1 K on a home made set up \cite{AC}. Specific heat measurement was performed by using the relaxation method at the temperature range from 0.06 K to 30 K on PPMS (Quantum Design) equipped with a $^3$He refrigerator insert (0.4--30 K) and a dilution refrigerator insert (0.06--0.4 K). The thermal conductivity ($\kappa$) was measured at temperatures down to 0.3 K and in magnetic fields up to 14 T in a $^3$He refrigerator by using a ``one heater, two thermometers" technique \cite{kappa-1, kappa-2, kappa-3}. In view of the crystal size, the heat current is always applied along the [110] direction throughout this work and the magnetic field is applied either along the [110] or [111] direction.

\section{Results and Discussion}

\subsection{Crystal Structure}

\begin{figure}
\centering\includegraphics[clip,width=8.5cm]{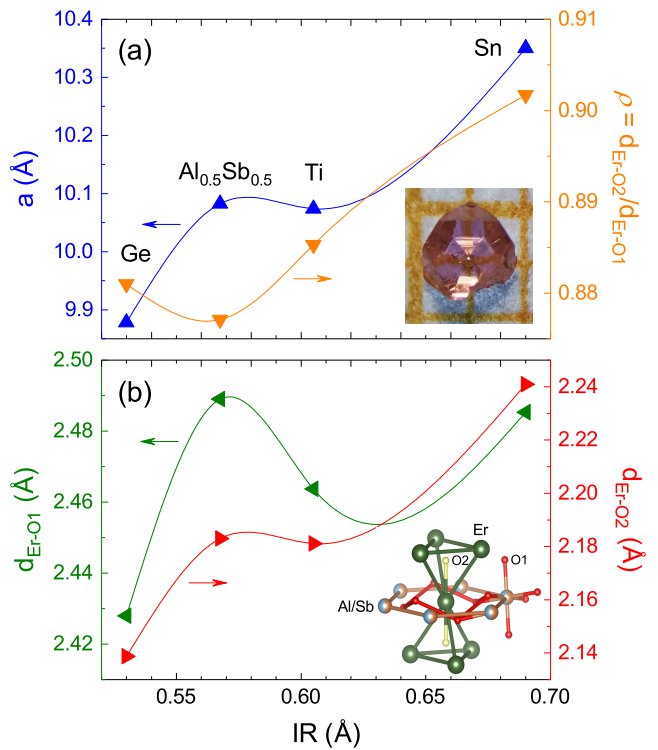}
\caption{(Color online) (a) Lattice constant and the Er-O bond length ratio vs ionic radius for the series of Er$_2$$B$$_2$O$_7$. Insert shows the photograph of a piece of Er$_2$AlSbO$_7$ single crystal. (b) Length of Er-O1 and Er-O2 bonds in an ErO$_8$ polyhedra for different Er$_2$$B$$_2$O$_7$ members. Inset is the local environment of a Er$^{3+}$ ion.}
\label{Structure}
\end{figure}

XRD results demonstrate that Er$_2$AlSbO$_7$ crystallizes in a cubic pyrochlore structure with $Fd\bar{3}m$ space group. Erbium sits at the 16$d$ (0.5, 0.5, 0.5), and both Al and Sb atoms locate at 16$c$ (0, 0, 0) with an equal concentration in a random distribution. The refined structural parameters are given in Table I and the comparison together with other Er$_2$$B$$_2$O$_7$ ($B$ = Ge, Ti, Sn) is plotted in Fig. \ref{Structure}. It should be mentioned that there is no clear relation between the magnetic ground state and the ionic radius (IR) of B site. Er$_2$Pt$_2$O$_7$ is not considered here since Pt$^{4+}$ ions with 5$d^6$ configuration does not have a closed $d$ electron shell configuration, which is rather different from the situations of empty $d$ orbitals for Ti$^{4+}$ or full-filled $d$ orbitals for Ge$^{4+}$ and Sn$^{4+}$. Indeed, the local coordination of Er$^{3+}$ ions and the distortion of ErO$_8$ polyhedra in Er$_2$Pt$_2$O$_7$ is quite unusual, which is discussed to be related to the spatially more extended 5$d$ orbitals of Pt$^{4+}$ ions \cite{Er2Pt2O7-1}. In Er$_2B_2$O$_7$ ($B$ = Ge, Ti, Sn) series, the lattice constant $a$ is almost linearly increased with increasing IR of $B^{4+}$ ion \cite{IR}, which implies a progressive decrement of chemical pressure on Er$^{3+}$ sublattice. However, the lattice constant of Er$_2$AlSbO$_7$ (10.0828 {\AA}) is found to be very close to that of Er$_2$Ti$_2$O$_7$ (10.074 {\AA}) \cite{Er2Ti2O7-XRD}. On account of the smaller average IR for Al$_{0.5}$/Sb$_{0.5}$ site, the comparable chemical pressure in Er$_2$AlSbO$_7$ implies a stronger distortion of ErO$_8$ polyhedra as compared with Er$_2$Ti$_2$O$_7$. It should be noted that the average IR for Er$_2$AlSbO$_7$ is a rough parameter to compare with other Er-based pyrochlores due to the charge misbalance of Al$^{3+}$ and Sb$^{5+}$ ions.

\begin{table}[!ht]
\caption{Room temperature crystallographic data for Er$_2$AlSbO$_7$.}
\begin{ruledtabular}
\begin{tabular}{ll}
Formula & Er$_2$AlSbO$_7$ \\
Formula weight & 595.25 \\
Wavelength ({\AA}) & 0.71 \\
Crystal system & Cubic \\
Space group & $Fd\bar{3}m$ \\
$a$ ({\AA}) & 10.0828(2) \\
$b$ ({\AA}) & 10.0828(2) \\
$c$ ({\AA}) & 10.0828(2) \\
$\alpha$ ($^{\circ}$) & 90 \\
$\beta$ ($^{\circ}$) & 90 \\
$\gamma$ ($^{\circ}$) & 90 \\
$V$ ({\AA}$^3$) & 1025.05(4) \\
$Z$ & 8 \\
Density (g cm$^{-3}$) & 7.714 \\
$\mu$ (mm$^{-1}$) & 37.834 \\
$R_1$, $wR_2$ ($F\rm_o >$ 4$\sigma$$F\rm_o$) & 0.0169, 0.0480 \\
$R_1$, $wR_2$ (all data) & 0.0173, 0.0482 \\
Goodness-of-Fit & 1.252 \\
\\
$x$ of O1 at 48$f$ ($x$, 1/8, 1/8) & 0.3277(3) \\
Er-O1 ($\times$ 6) ({\AA}) & 2.489(3) \\
Er-O2 ($\times$ 2) ({\AA}) & 2.18299(5) \\
$\rho$ [$\equiv$$d$$\rm_{Er-O2}$/$d$$\rm_{Er-O1}$] & 0.8771 \\
\end{tabular}
\end{ruledtabular}
\begin{tablenotes}
\footnotesize
\item $R_1=\Sigma\mid\mid F{\rm_o}\mid-\mid F{\rm_c}\mid\mid/\Sigma\mid F{\rm_o}\mid$, $wR_2=[\Sigma w({F{\rm_o}}^2-{F{\rm_c}}^2)^2/\Sigma w({F{\rm_o}}^2)^2]^{1/2}$
\end{tablenotes}
\end{table}

The distortion degree of ErO$_8$ polyhedra can be parameterized using the ratio $\rho$$\equiv$$d$$\rm_{Er-O2}$/$d$$\rm_{Er-O1}$ to characterize the axial distortion along the local [111] direction \cite{Er2Ge2O7-1}. As displayed in Fig. \ref{Structure}(a), Er$_2$AlSbO$_7$ has the smallest $\rho$ which demonstrates the strongest axial distortion among Er$_2B_2$O$_7$ series. The unusually small $\rho$ in Er$_2$AlSbO$_7$ is associated with the much longer Er-O1 bonds as shown in Fig. \ref{Structure}(b). It is clearly seen from the inset to Fig. \ref{Structure}(b) that the nonmagnetic ions are only coordinated with O1 atoms, and the Al/Sb disorder is probably responsible for the abnormally longer Er-O1 in Er$_2$AlSbO$_7$. On the other hand, the charge misbalance of the Al/Sb occupancy can also give rise to a different local CEF environment of Er$^{3+}$ ions via the shared O1 atoms. It seems that charge mixture can efficiently modify CEF without altering the average lattice constant much, which probably provides a feasible route to search for interesting phenomena in the already known pyrochlores.

\subsection{Magnetic Susceptibility}

\begin{figure*}
\centering\includegraphics[clip,width=16cm]{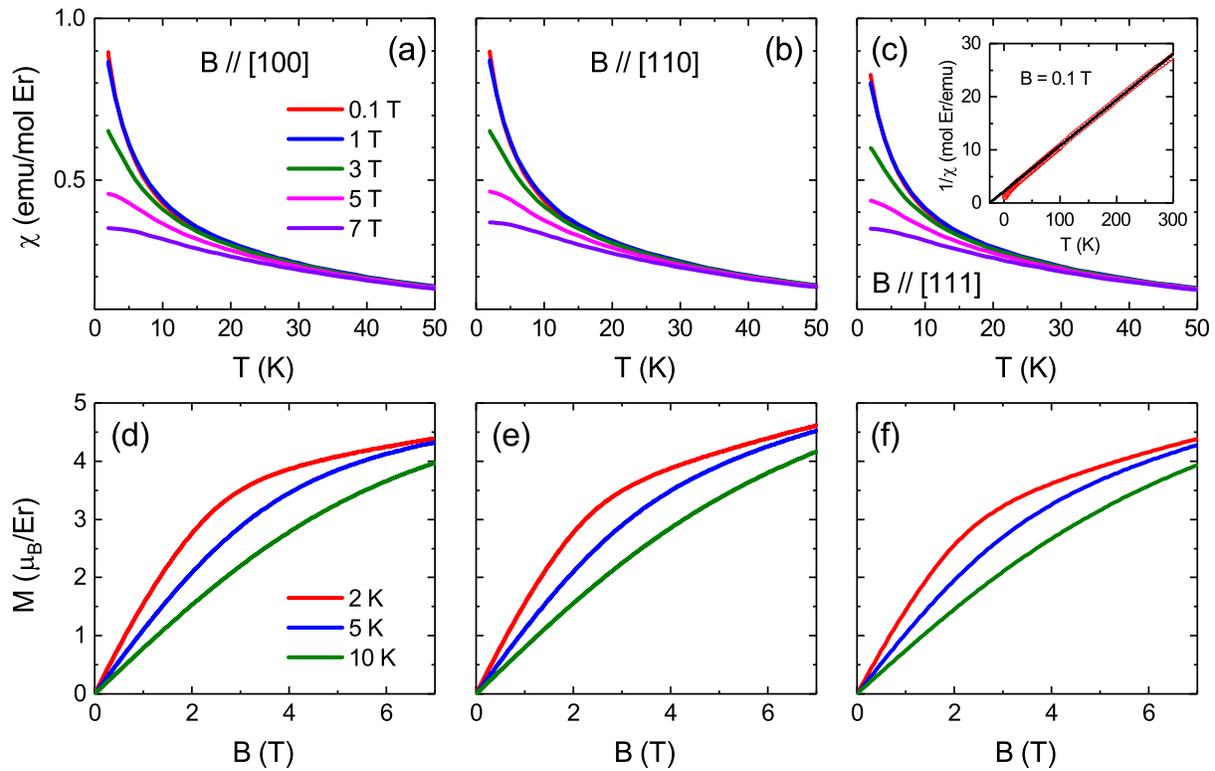}
\caption{(Color online) (a-c) Temperature dependencies of the magnetic susceptibility measured in fields from 0.1 T to 7 T along three directions. (d-f) Magnetization curves measured at 2 K, 5 K, and 10 K along three directions. Inset to panel (c) is the Curie-Weiss fit for data measured in 0.1 T.}
\label{DC}
\end{figure*}

Figure \ref{DC} shows the magnetic susceptibility $\chi (T)$ and magnetization curves along three characteristic axes [100], [110], and [111]. Although Er$^{3+}$ spins are highly anisotropic, the macroscopic magnetic properties look isotropic and are similar to those of Er$_2$Ti$_2$O$_7$ \cite{Er2Ti2O7-10,Er2Ti2O7-crystal growth}. $\chi (T)$ is increased upon lowering temperature and no anomaly associated with a long-range magnetic order is observed down to 2 K. The high-temperature $\chi (T)$ follows well the Curie-Weiss law and the fit between 100 and 300 K along the [111] direction measured in 0.1 T yields a Weiss temperature $\theta\rm_{CW}$ = -25.0(4) K and an effective moment $\mu\rm_{eff}$ = 9.62 $\mu\rm_B$/Er$^{3+}$. The obtained $|\theta\rm_{CW}|$ is close to that of Er$_2$Ge$_2$O$_7$ \cite{Er2Ge2O7-1} but a little larger than that of Er$_2$Ti$_2$O$_7$ \cite{Er2Ti2O7-6,Er2Ti2O7-9} and Er$_2$Sn$_2$O$_7$ \cite{Er2Sn2O7-3,Er2Sn2O7-4}. The deduced $\mu\rm_{eff}$ is comparable to that for other Er$_2B_2$O$_7$ \cite{Er2Ti2O7-6,Er2Ti2O7-9,Er2Ge2O7-1,Er2Sn2O7-3,Er2Sn2O7-4}, and is consistent with the expectation of 9.58 $\mu\rm_B$ for the $^4I_{15/2}$ ground state of Er$^{3+}$ ions. As increasing field, the moment is monotonously increased and achieves to about 4.5 $\mu\rm_B$/Er$^{3+}$ in 7 T. No evidence for a field-induced magnetic transition is found from the derivative of the magnetization curve (not shown).

\begin{figure}
\centering\includegraphics[clip,width=6.0cm]{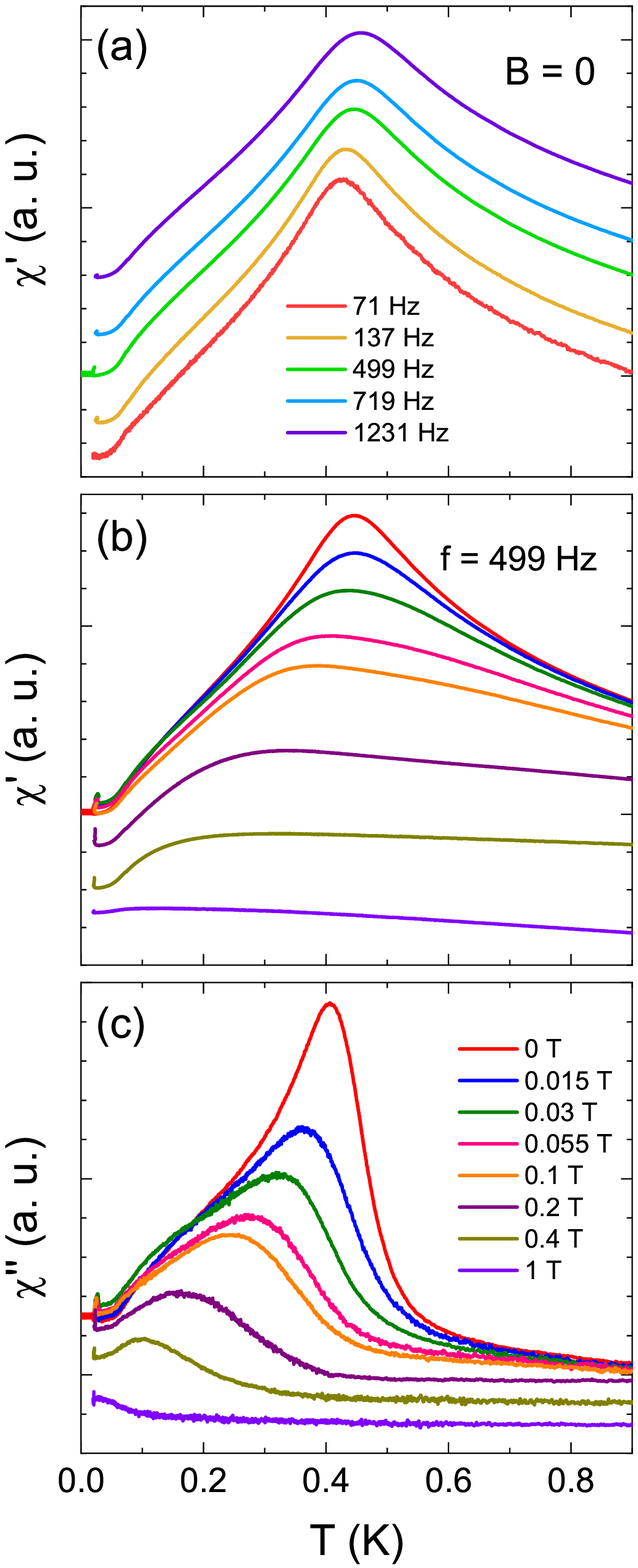}
\caption{(Color online) Temperature dependencies of the ac magnetic susceptibility measured in zero field with different frequencies (a) and in different dc fields with $f$ = 499 Hz (b-c).}
\label{AC-1}
\end{figure}

To further characterize the magnetic ground state of Er$_2$AlSbO$_7$, ac susceptibility was measured down to 20 mK. Figure \ref{AC-1}(a) shows the temperature dependence of the real part $\chi'$ measured in the absence of dc field. Clearly, a frequency-dependent peak is observed which is a typical dynamic behavior for a spin-glass-like system. The $\chi'$ and the imaginary part $\chi''$ for $f$ = 499 Hz as a function of temperature in different dc fields are shown in Figs. \ref{AC-1}(b,c). As the field is increased, the peak position is shifted to lower temperatures and the magnitudes of both $\chi'$ and $\chi''$ are suppressed. In 1 T, the peak is almost smeared out.

\begin{figure}
\centering\includegraphics[clip,width=6.5cm]{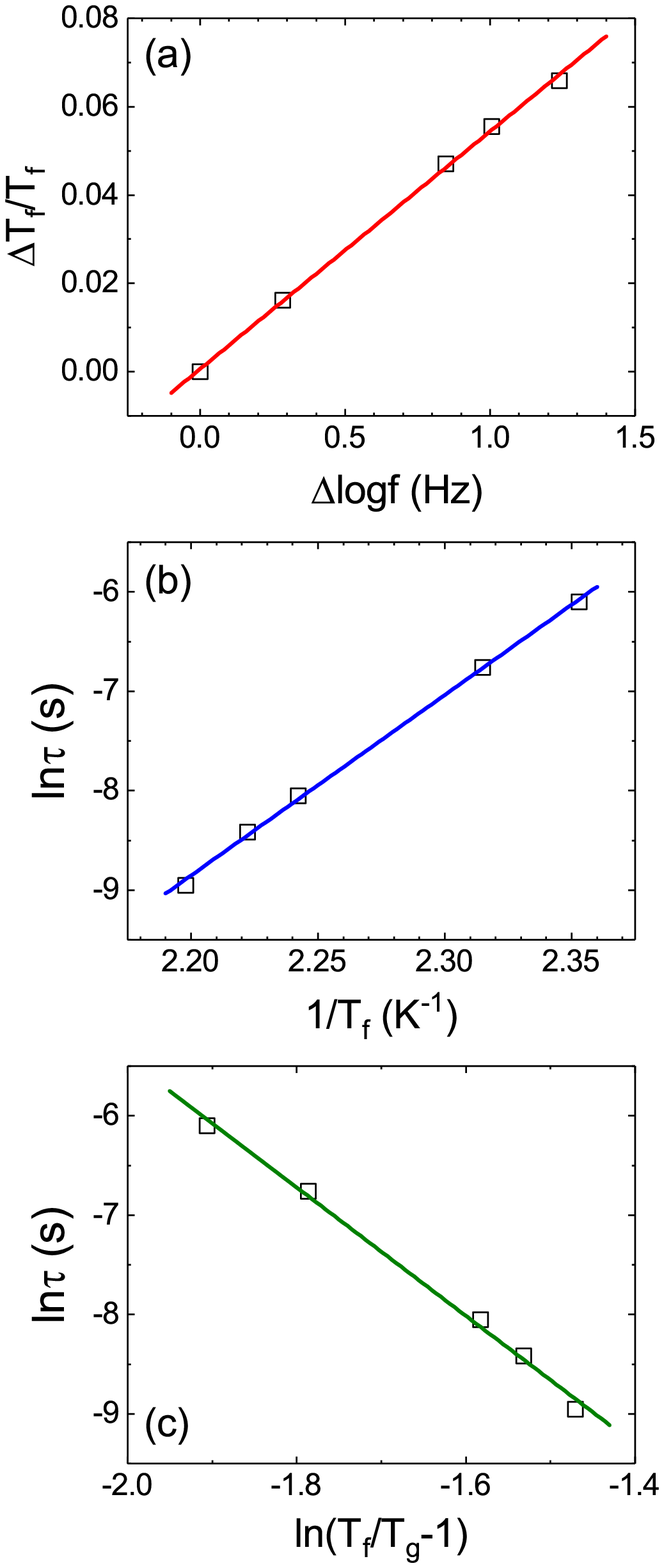}
\caption{(Color online) (a) Frequency dependence of the freezing temperature plotted as $\Delta T_f$/$T_f$ versus $\Delta$log$f$. The slope of the fitted straight line is the Mydosh parameter $\delta$. (b-c) Arrhenius law and power-law fits for the temperature dependence of the relaxation time $\tau$.}
\label{AC-2}
\end{figure}

A phenomenological classification can be done by estimating the Mydosh parameter $\delta$ = $\Delta T_f$/($T_f$$\Delta$log$f$) \cite{Mydosh-1, Mydosh-2}, where $T_f$ is the freezing temperature defined as the peak position in $\chi'(T)$. This is a quantitative measure of the relative peak shift in $\chi'$ with frequency and usually used as a criterion to distinguish the conventional spin-glass, cluster-glass, and superparamagnetic systems. For Er$_2$AlSbO$_7$, $\delta$ is calculated to be 0.054(1) as seen in Fig. \ref{AC-2}(a). This value is intermediate between those reported for the canonical spin-glass systems ($\delta \sim$ 0.005) and those for noninteracting ideal superparamagnetic systems ($\delta \sim$ 0.1) \cite{Mydosh-1, Mydosh-2}. The dynamic behavior in Er$_2$AlSbO$_7$ is an intermediate situation, likely the so-called cluster glass. Indeed, the $\delta$ value is comparable to those in the known cluster-glass compounds, e.g., 0.086 for PrRhSn$_3$ \cite{PrRhSn3} and 0.062 for CeNi$_{0.9}$Cu$_{0.1}$ \cite{CeNiCu}. The name ``cluster glass'' or ``mictomagnetism'' was firstly proposed in the metallurgy with high concentration of impurities in alloys, which can be considered as an ensemble of interacting magnetic clusters. At low temperatures, these clusters will freeze with random orientation in a manner analogous to the spin-glass freezing \cite{Mydosh-1}. The random distribution of Al$^{3+}$ and Sb$^{5+}$ probably acts as a role of highly concentrated impurities and hinders the development of the magnetic order in Er$_2$AlSbO$_7$.

Next, we analyze the frequency dependence of the freezing temperature $T_f$. Arrhenius law is usually used to describe the thermally activated processes for a superparamgnet,
\begin{equation}
\tau = \tau_0{\rm exp}(E_a/T_f).
\end{equation}
Here, $\tau$ = 1/$2\pi f$ is the relaxation time and $E_a$ is the activation energy barrier. The linear fit in Fig. \ref{AC-2}(b) gives $E_a$ = 18.2(4) K and $\tau_0 \sim 10^{-22}$ s. It is known that for the characteristic time $\tau_0$ is about $10^{-9} \sim 10^{-13}$ s in a noninteracting superparamagnetic system \cite{Mydosh-2}. The obtained extremely short $\tau_0$ is thus unphysical and implies that additional interaction among spins is present in the system.

Vogel-Fulcher law is a modification of the Arrhenius law which takes the interactions among the spin clusters into account,
\begin{equation}
\tau = \tau_0{\rm exp}[E_a/(T_f-T_0)],
\end{equation}
where $T_0$ is the Vogel-Fulcher temperature and is a measure of the inter-cluster interaction strength. According to the equation, a linear relation is expected when plotting ln$\tau$ versus 1/($T_f - T_0$) for an appropriate $T_0$. It is found that the time $\tau_0$ becomes reasonable with adding the $T_0$ term. However, the yielded $E_a$ and $\tau_0$ are very sensitive to the selected $T_0$ (not shown). Nevertheless, the necessity of $T_0$ supports the presence of the interactions among magnetic clusters.

Another approach is the critical scaling formula assuming that there exists a true equilibrium phase transition at $T\rm_g$,
\begin{equation}
\tau = \tau_0(T_f/T{\rm_g} - 1)^{-zv}.
\end{equation}
In the vicinity of the phase transition, the relaxation time $\tau$ follows a power-law divergence of critical slowing down with $zv$ the dynamic critical exponent. With three parameters ($\tau_0$, $T\rm_g$, $zv$), a reasonable fit is shown in Fig. \ref{AC-2}(c) with a fixed $T\rm_g$ = 0.37 K. The obtained $zv$ = 6.5(2) is well within the range 4 $< zv <$ 12 for different spin-glass systems \cite{Mydosh-1,zv-1,zv-2}, and the characteristic relaxation time $\tau_0 \sim 10^{-8}$ s is consistent with those reported in cluster-glass compounds.

\subsection{Specific Heat}

\begin{figure}
\centering\includegraphics[clip,width=8.5cm]{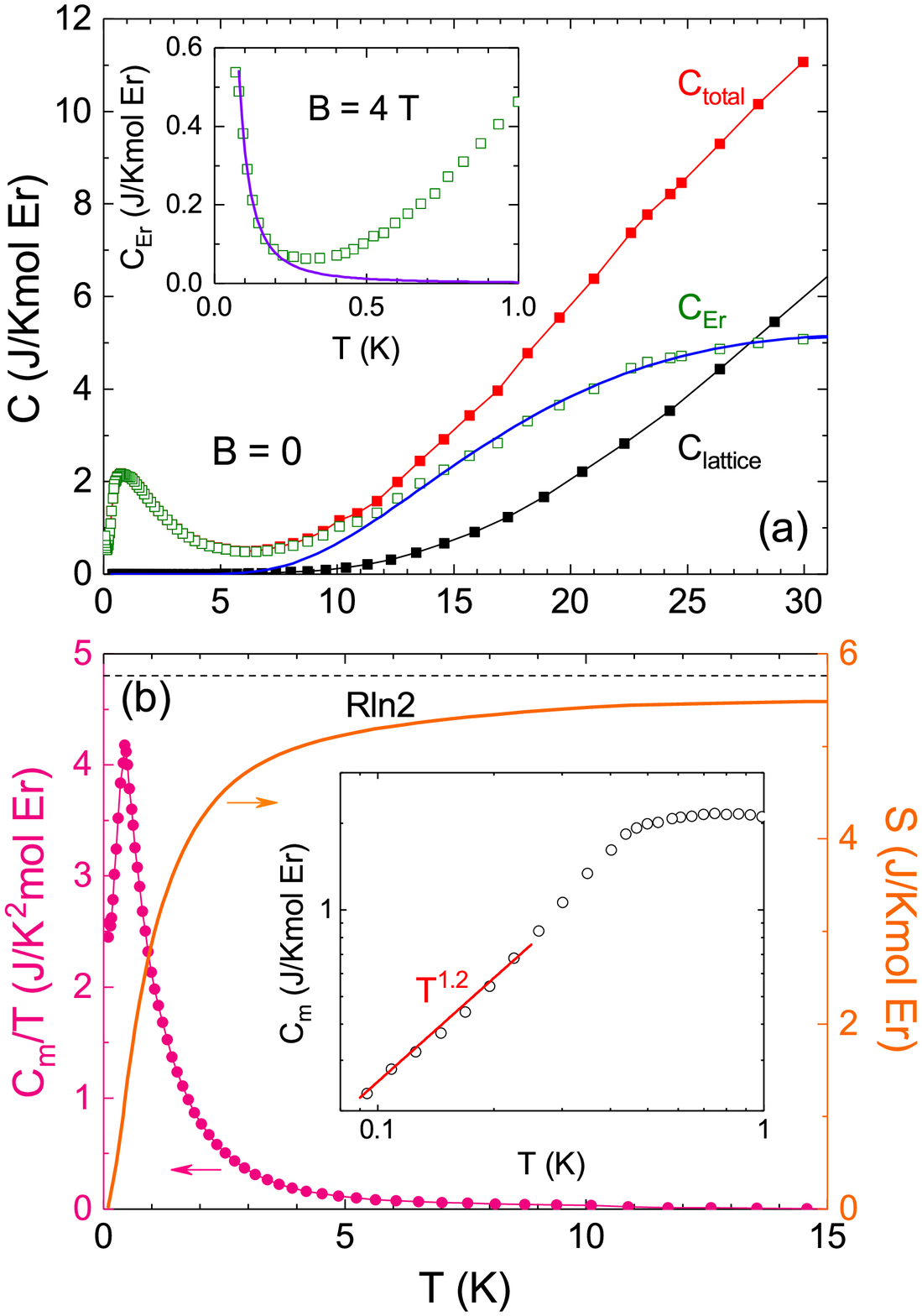}
\caption{(Color online) (a) Temperature dependence of the specific heat measured in the absence of magnetic field. Inset is the $C\rm_{Er}$ meaured in 4 T below 1 K. The blue and purple solid lines are the fitted curves of the CEF and nuclear contributions, respectively. (b) The magnetic specific heat and the calculated entropy. The dashed line indicates the expected value of $R$ln2 for $S$ = 1/2 spin systems. Inset is the zoom in of $C\rm_m$ below 1 K, which follows a $T^{1.2}$ dependence below 0.25 K.}
\label{Cp-1}
\end{figure}

Figure \ref{Cp-1}(a) shows the temperature dependence of the specific heat $C\rm_{total}$ measured between 60 mK and 30 K and in the absence of the magnetic field. No $\lambda$ anomaly is observed, which is in agreement with the dc and ac susceptibility and is a further support for the absence of magnetic order. Instead, a broad maximum is found around 0.8 K. Similar behavior is also observed in the isostructural Er$_2$GaSbO$_7$ \cite{Er2GaSbO7}. In order to get more information about the magnetic specific heat $C\rm_m$ and estimate the entropy change, the specific heat of the isostructural nonmagnetic compound Lu$_2$AlSbO$_7$ is measured and used as the lattice reference $C\rm_{lattice}$, as shown in Fig. \ref{Cp-1}(a). After subtracting $C\rm_{lattice}$, the contribution from the magnetic Er$^{3+}$ ions, named as $C\rm_{Er}$, is obtained. As seen in Fig. \ref{Cp-1}(a), a broad hump of $C\rm_{Er}$ is present above 15 K, which is likely originated from the Er$^{3+}$ CEF levels and is named $C\rm_{CEF}$, similar to the case in Er$_2$Ge$_2$O$_7$ \cite{Er2Ge2O7-1}. In the $D_{3d}$ symmetry, the 16-fold degenerate Er$^{3+}$ ion with $J$ = 15/2 split into eight doublets. In this work, the $C\rm_{CEF}$ contribution in Er$_2$AlSbO$_7$ is simulated including the first two doublets using the following expression,
\begin{equation}
C{\rm_{CEF}} = R\frac{(E_1/T)^2g_0g_1e^{-E_1/T} + (E_2/T)^2g_0g_2e^{-E_2/T}}{(g_0 + g_1e^{-E_1/T} + g_2e^{-E_2/T})^2}.
\end{equation}
Here, $g_0$ = $g_1$ = $g_2$ = 2 are the degeneracies for the three doublets, $E_1$ and $E_2$ are the energy gaps of the first and second excited doublets separated from the ground state, and $R$ is the universal gas constant. As can be seen in Fig. \ref{Cp-1}(a) (blue line), the high-temperature hump feature is well reproduced between 15 and 30 K, yielding $E_1$ = 62 K and $E_2$ = 123 K. The energy of the first excited doublet is comparable to those reported in other Er$_2$$B_2$O$_7$ compounds, determined by inelastic neutron scattering measurements \cite{Review-XY pyrochlores-3}, i.e., 76 K for Ge, 73 K for Ti, 58 K for Sn, while the second excited level is a little higher than that for Ti (85 K) and Sn (86 K) but close to Ge (108 K).

Besides the CEF contribution, there is also an upturn below 0.14 K which is ascribed to the nuclear contribution of Er$^{3+}$ ions. The expression is as following \cite{Er2Ge2O7-2},
\begin{equation}
C{\rm_n} = \frac{R}{T^2}\frac{\sum\limits_{i=-I}^{i=I}\sum\limits_{j=-I}^{j=I}(W_i^2 - W_iW_j)e^{-\frac{W_i+W_j}{T}}}{\sum\limits_{i=-I}^{i=I}\sum\limits_{j=-I}^{j=I}e^{-\frac{W_i+W_j}{T}}},
\end{equation}
where $I$ is the nuclear spin and the energy levels $W_i$ are given by
\begin{equation}
W_i = -a'i + P(i^2 - \frac{1}{3I(I + 1)}),
\end{equation}
\begin{equation}
i = -I, -I + 1, ..., I - 1, I.
\end{equation}
Here, $a'$ is the magnetic interaction parameter and $P$ is the quadrupole coupling constant. With $I$ = 7/2, the $C\rm_n$ contribution is extracted with $a'$ and $P$ as adjustable parameters. Although the nuclear specific heat is almost independent of the magnetic field, it is notable that the upturn behavior is a bit field dependent, which is likely ascribed to the contribution from the unrecovered electrons at low temperatures. To get more accurate $C\rm_m$, we analyze the $C\rm_n$ data measured in 4 T with a sharper upturn. As shown in the inset to Fig. \ref{Cp-1}(a), the best fit gives $a' =$ 0.007 K and $P =$ -0.002 K.

\begin{figure}
\centering\includegraphics[clip,width=8cm]{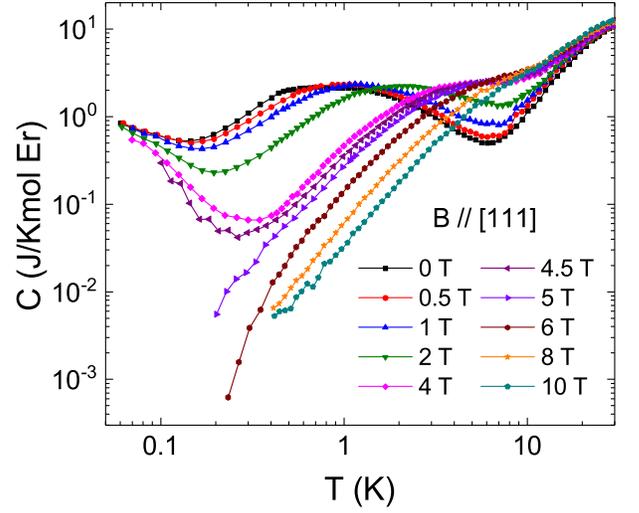}
\caption{(Color online) (a) Temperature dependencies of the specific heat measured in the [111] direction field up to 10 T.}
\label{Cp-2}
\end{figure}

Finally, the magnetic contribution of Er$^{3+}$ ions, $C\rm_m$, is obtained by subtracting $C\rm_{lattice}$, $C\rm_{CEF}$, and $C\rm_n$ from $C\rm_{total}$, as shown in Fig. \ref{Cp-1}(b). Below 0.25 K, $C\rm_m$ follows a $T^{1.2}$ dependence, as seen in the inset to Fig. \ref{Cp-1}(b), which is a further evidence for the spin-freezing state in Er$_2$AlSbO$_7$ \cite{Mydosh-1,Mydosh-2}. The recovered magnetic entropy is about 5.5 J/Kmol, which approaches the expected value of $R$ln2 for $S$ = 1/2 spin systems. This implies that Er$^{3+}$ ions in Er$_2$AlSbO$_7$ can also be described as an effective spin $S =$ 1/2.

The measured specific heat $C\rm_{total}$ under the [111] direction field are given in Fig. \ref{Cp-2}. With increasing field, the broad peak around 0.8 K observed in zero field moves to higher temperatures and almost disappears in 10 T. The field dependence of this broad peak, however, is not likely ascribed to the Schottky anomaly since the first excited doublet is about 60 K higher than the ground state doublet. In addition, the magnetic entropy associated with this broad peak is almost recovered to $R$ln2. Therefore, the broad peak around 1 K is likely related to the spin degrees of freedom within the isolated ground state doublet \cite{Review-XY pyrochlores-1}.

\subsection{Thermal Conductivity}

\begin{figure}
\centering\includegraphics[clip,width=7.5cm]{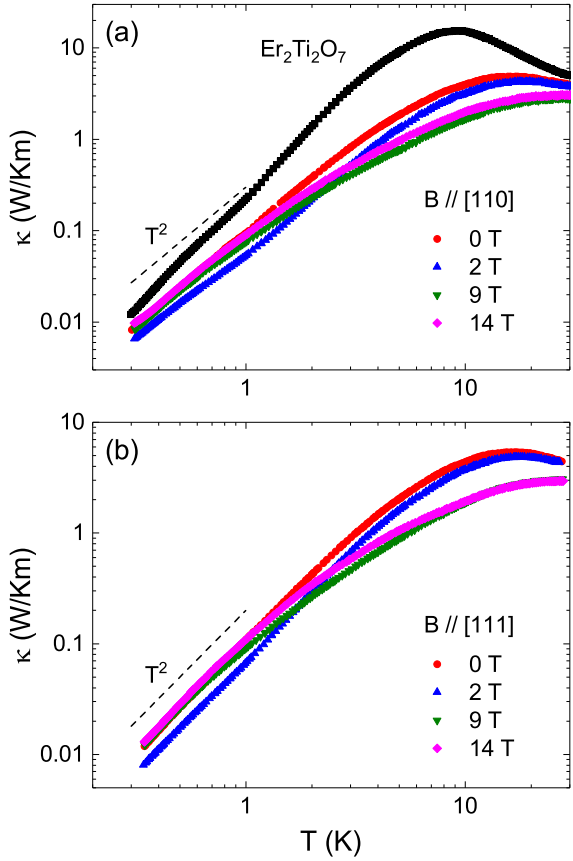}
\caption{(Color online) Temperature dependencies of the thermal conductivities of Er$_2$AlSbO$_7$ single crystal measured in the magnetic field along the (a) [110] and (b) [111] direction, respectively. For comparison, the zero-field thermal conductivity data for Er$_2$Ti$_2$O$_7$, taken from Ref. \onlinecite{kappa-1}, are also shown in panel (a). The dashed lines indicate the $T^2$ dependence.}
\label{kappaT}
\end{figure}

Temperature dependencies of the thermal conductivity with applied fields along the [110] and [111] directions are displayed in Fig. \ref{kappaT}. The behavior for both field directions are almost the same. It is known that in the magnetic insulators, the magnetic phase transitions can lead to clear anomalies in the $\kappa(T)$ or $\kappa(B)$ curves, due to the phonon scattering by magnetic critical fluctuations \cite{kappa-1, kappa-2, kappa-3}. In zero field, the $\kappa(T)$ curves of Er$_2$AlSbO$_7$ single crystal exhibit smooth temperature dependence, in good consistence with the absence of a magnetic order. In addition, a very weak phonon peak is observed around 16 K with the magnitude about 5 W/Km. This is in sharp contrast to the case of Er$_2$Ti$_2$O$_7$, in which a clear peak is present around 8 K and the magnitude is three times larger ($\sim$ 15 W/Km) \cite{kappa-1}. It is likely due to the strong lattice disorder caused by the random Al/Sb occupancy. Below 1 K, the $\kappa$ nearly follows a $T^2$ dependence, which is strongly deviated from the expected $T^3$ dependence in the boundary scattering limit \cite{Berman}. Since a spin-freezing transition of Er$^{3+}$ ions is developed at lower temperatures, this phenomenon may suggest phonon scattering caused by spin fluctuations in Er$_2$AlSbO$_7$ \cite{kappa-1, kappa-2, kappa-3}. However, the possibilities of specular reflection effect and disorder-induced decrease of the mean free path of phonons, which may also result in a $T^2$ dependence, can not be excluded \cite{specular-reflection}. When applying a field of 2 T, the $\kappa$ becomes much smaller at low temperatures, and a slight slope change is detected around 1.5 K. In higher fields, the $\kappa$ is recovered to the zero-field value and also follows the $T^2$ dependence below 1 K, demonstrating that the spin fluctuations are still active even in 14 T.

\begin{figure}
\centering\includegraphics[clip,width=7.5cm]{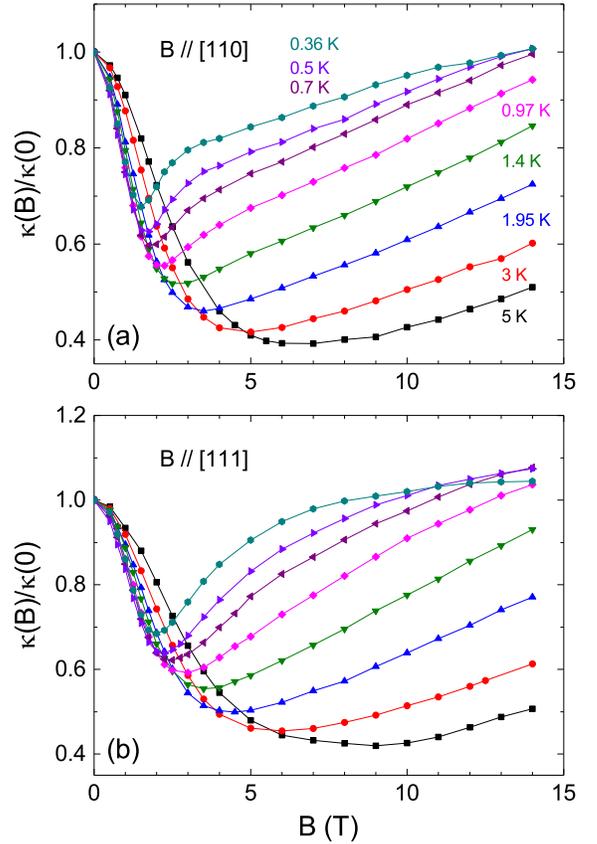}
\caption{(Color online) Thermal conductivity as a function of the magnetic field along the (a) [110] and (b) [111] directions.}
\label{kappaH}
\end{figure}

Figure \ref{kappaH} shows the field dependencies of the thermal conductivity under the field along the [110] and [111] directions. For both directions, the $\kappa(B)$ isotherms exhibit a broad dip-like feature, and the dip position shifts to higher fields upon warming. The nearly isotropic behaviors resemble the paramagnetic scattering of phonons \cite{kappa-4, kappa-5}. However, these $\kappa(B)$ curves cannot be well fitted by using a simple paramagnetic scattering formula, which yields much narrower dip structure (not shown in the figure) \cite{kappa-4, kappa-5}. In addition, a slight curvature is visible around 3 T below 0.7 K for $B$ $\parallel$ [110], while the $\kappa(B)$ curves are rather smooth for $B$ $\parallel$ [111]. Therefore, it is more likely that the strong field dependence of $\kappa$ is due to the phonon scattering by spin fluctuations, which can only be suppressed in very high magnetic field. In this regard, it is worthy of noting that in another rare-earth titanate Tb$_2$Ti$_2$O$_7$ the phonon scattering by spin fluctuations are extremely strong, which results in a phonon-glass-like behavior with very short phonon mean free path \cite{kappa-6}.

\section{Discussions}

To summarize the main experimental results of Er$_2$AlSbO$_7$, no signature for the long-range magnetic order is observed down to 20 mK. Instead, a spin-freezing ground state is possibly developed, judged from the frequency dependence of the ac susceptibility and the $T^{1.2}$ dependence of the magnetic specific heat. The low-temperature thermal conductivity data suggest the existence of rather strong spin fluctuations.

According to the frequency dependence of the peak position in the ac susceptibility results, a cluster spin-glass state is likely realized in Er$_2$AlSbO$_7$. In Er-based pyrochlores, Er$_2$Ti$_2$O$_7$ is a fascinating compound due to the novel ground-state selection via order-by-disorder mechanisms. It is widely accepted that the $\psi_2$ structure is stabilized through zero-point quantum fluctuation \cite{Er2Ti2O7-2}. However, two theoretical works predicted that the structural disorder in Er$_{2-x}$Y$_x$Ti$_2$O$_7$ prefer to select the $\psi_3$ state for small $x$ \cite{Er2-xYxTi2O7-theory-1, Er2-xYxTi2O7-theory-2}. Neutron scattering measurements have indeed observed the collapse of the spin gap which is a direct measure of the selection of $\psi_2$ over $\psi_3$ state, and suggested that the ground state for 20\%-Y doping sample is a frozen mosaic of $\psi_2$ over $\psi_3$ domains \cite{Er2-xYxTi2O7-experiment-2}. Recently, a cluster spin-glass state is predicted to be stabilized at a stronger disorder \cite{CGS}. In Er$_2$AlSbO$_7$, the random distribution of Al and Sb atoms can be considered as a kind of exchange disorder which might be strong enough to stabilize a cluster spin-glass ground state. One may note that a sister material Er$_2$GaSbO$_7$ has been also synthesized \cite{Er2GaSbO7}. It is apparently has weaker randomness than Er$_2$AlSbO$_7$. However, since the ground state of Er$_2$GaSbO$_7$ is still unknown \cite{Er2GaSbO7}, one cannot compare the magnetic properties and disorder effect between Er$_2$GaSbO$_7$ and Er$_2$AlSbO$_7$.

In fact, the cluster spin-glass state driven by the exchange disorder has been observed in Na$A'$Co$_2$F$_7$ ($A'$ = Ca, Sr), two recently discovered $3d$ transition metal pyrochlore fluorides with the randomly distributed Na$^+$ and Ca$^{2+}$/Sr$^{2+}$ ions. Magnetic susceptibility measurements suggested that Co$^{2+}$ ions in both compounds undergo a spin-freezing transition at low temperatures \cite{NaCaCo2F7-1, NaSrCo2F7-1}. A magnetic model consisted of short-ranged $XY$ AFM clusters was proposed as the ground state \cite{NaCaCo2F7-2, NaCaCo2F7-3}, which are built from the linear combination of $\psi_2$ and $\psi_3$.

Besides the inherent structural disorder, the phase competition between different ordered states in the $XY$ anisotropic exchange phase diagram might be also responsible for the significant suppression or even disappearance of the magnetic order. The presence of the broad anomaly in the specific heat below 10 K is conjectured to be an experimental clue of phase competition in Er-based pyrochlores \cite{Review-XY pyrochlores-1}. In Er$_2$$B_2$O$_7$ series, only Er$_2$Ti$_2$O$_7$ and Er$_2$Ge$_2$O$_7$ order above 1 K and do not display such a broad anomaly, which is consistent with the phase diagram that Er$_2$Ti$_2$O$_7$ locates deeply inside the $\Gamma_5$ phase and far away from the phase boundary. The robust $\Gamma_5$ state to impurities in Er$_{2-x}$Y$_x$Ti$_2$O$_7$ also confirms that Er$_2$Ti$_2$O$_7$ is far-removed from competing $XY$ phases. In contrast, Er$_2$Pt$_2$O$_7$ and Er$_2$Sn$_2$O$_7$ with reduced $T\rm_N$ and stronger $XY$ anisotropy display a broad anomaly above $T\rm_N$. It has been confirmed from the inelastic neutron measurement that the presence of the broad anomaly in Er$_2$Pt$_2$O$_7$ is resulted from the short-range spin fluctuations, which is originated from the competition between $\Gamma_5$ and $\Gamma_7$ phases \cite{Er2Pt2O7-2}. Er$_2$Sn$_2$O$_7$, however, locates in close proximity to the boundary between $\Gamma_5$ and $\Gamma_7$ phases in the anisotropic exchange phase diagram also implies a significant phase competition \cite{Er2Sn2O7-2}. In Er$_2$AlSbO$_7$, the broad anomaly moves to higher temperatures with increasing field, in opposite to the tendency that the freezing temperature is gradually suppressed as seen in the ac susceptibility measurement. This implies that the broad anomaly in Er$_2$AlSbO$_7$ is probably resulted from the phase competition and the magnetic order is destroyed by the significant spin fluctuations. The thermal conductivity data also suggest that there are rather strong spin fluctuations that can scatter phonons. However, since both the spin configuration and the anisotropic exchange interaction are not known at present, it is not clear which phases are involved into the phase competition. The neutron scattering measurements are called for clarifying this issue.

\section{Summary}

Single crystals of a new rare-earth pyrochlore Er$_2$AlSbO$_7$, in which Al$^{3+}$ and Sb$^{5+}$ atoms share the same positions with a random distribution, were grown by flux method. Different from the other reported Er-based pyrochlores, no signature for the long-range magnetic order is observed in Er$_2$AlSbO$_7$ at temperatures down to several tens of milli-kelvin. Instead, a spin-freezing transition is detected below 0.37 K, which is primarily ascribed to the inherent structural disorder. A cluster spin-glass state is likely formed based on the frequency dependence of the peak position in the ac susceptibility. The temperature and field dependence of thermal conductivity indicates rather strong spin fluctuations which is probably resulted from the phase competition. It seems that charge mixture can efficiently modify CEF without altering the average lattice constant much, which probably provides a feasible route to search for interesting phenomena in the already known pyrochlores.

\begin{acknowledgements}

This work was supported by the National Natural Science Foundation of China (Grants No. U1832209, No. 51702320, No. U1832166, and No. 11874336), the National Basic Research Program of China (Grant No. 2016YFA0300103), the Innovative Program of Hefei Science Center CAS (Grant No. 2019HSC-CIP001), and the Users with Excellence Project of Hefei Science Center CAS (Grant No. 2018HSC-UE012). Q.H. and H.Z. thank the support from NSF-DMR under award NSF-DMR-1350002. A portion of this work was performed at the National High Magnetic Field Laboratory, which is supported by the National Science Foundation Cooperative Agreement No. DMR-1644779 and the State of Florida.

\end{acknowledgements}


\begin{thebibliography}{}

\bibitem{Review-1}
J. S. Gardner, M. J. P. Gingras, and J. E. Greedan, Rev. Mod. Phys. {\bf 82}, 53 (2010).

\bibitem{Review-XY pyrochlores-1}
A. M. Hallas, J. Gaudet, and B. D. Gaulin, Annu. Rev. Condens. Matt. Phys. {\bf 9}, 105 (2018).

\bibitem{Review-XY pyrochlores-2}
J. G. Rau and M. J. P. Gingras, Annu. Rev. Condens. Matt. Phys. {\bf 10}, 357 (2019).

\bibitem{Review-XY pyrochlores-3}
J. Gaudet, A. M. Hallas, A. I. Kolesnikov, and B. D. Gaulin, Phys. Rev. B {\bf 97}, 024415 (2018).

\bibitem{Review-XY pyrochlores-4}
J. S. Gardner, M. J. P. Gingras, and J. E. Greedan, Rev. Mod. Phys. {\bf 82}, 53 (2010).

\bibitem{Phase-diagram-1}
A. W. C. Wong, Z. Hao, and M. J. P. Gingras, Phys. Rev. B {\bf 88}, 144402 (2013).

\bibitem{Phase-diagram-2}
L. D. C. Jaubert, O. Benton, J. G. Rau, J. Oitmaa, R. R. P. Singh, N. Shannon, and M. J. P. Gingras, Phys. Rev. Lett. {\bf 115}, 267208 (2015).

\bibitem{Phase-diagram-3}
H. Yan, O. Benton, L. Jaubert, and N. Shannon, Phys. Rev. B {\bf 95}, 094422 (2017).

\bibitem{Er2Ti2O7-1}
R. Siddharthan, B. S. Shastry, A. P. Ramirez, A. Hayashi, R. J. Cava, and S. Rosenkranz, Phys. Rev. Lett. {\bf 83}, 1854 (1999).

\bibitem{Er2Ti2O7-2}
J. D. M. Champion, M. J. Harris, P. C. W. Holdsworth, A. S. Wills, G. Balakrishnan, S. T. Bramwell, E. \v{C}i\v{z}m\'{a}r, T. Fennell, J. S. Gardner, J. Lago, D. F. McMorrow, M. Orend\'{a}\v{c}, A. Orend\'{a}\v{c}ov\'{a}, D. M. Paul, R. I. Smith, M. T. F. Telling, and A. Wildes, Phys. Rev. B {\bf 68}, 020401(R) (2003).

\bibitem{Er2Ti2O7-3}
A. Poole, A. S. Wills, and E. Leli\`{e}vre-Berna, J. Phys.: Condens. Matt. {\bf 19}, 452201 (2007).

\bibitem{Er2Ti2O7-4}
J. P. C. Ruff, J. P. Clancy, A. Bourque, M. A. White, M. Ramazanoglu, J. S. Gardner, Y. Qiu, J. R. D. Copley, M. B. Johnson, H. A. Dabkowska, and B. D. Gaulin, Phys. Rev. Lett. {\bf 101}, 147205 (2008).

\bibitem{Er2Ti2O7-5}
S. S. Sosin, L. A. Prozorova, M. R. Lees, G. Balakrishnan, and O. A. Petrenko, Phys. Rev. B {\bf 82}, 094428 (2010).

\bibitem{Er2Ti2O7-6}
P. Dalmas de R\'{e}otier, A. Yaouanc, Y. Chapuis, S. H. Curnoe, B. Grenier, E. Ressouche, C. Marin, J. Lago, C. Baines, and S. R. Giblin, Phys. Rev. B {\bf 86}, 104424 (2012).

\bibitem{Er2Ge2O7-1}
X. Li, W. M. Li, K. Matsubayashi, Y. Sato, C. Q. Jin, Y. Uwatoko, T. Kawae, A. M. Hallas, C. R. Wiebe, A. M. Arevalo-Lopez, J. P. Attfield, J. S. Gardner, R. S. Freitas, H. D. Zhou, and J. G. Cheng, Phys. Rev. B {\bf 89}, 064409 (2014).

\bibitem{Er2Ge2O7-2}
Z. L. Dun, X. Li, R. S. Freitas, E. Arrighi, C. R. Dela Cruz, M. Lee, E. S. Choi, H. B. Cao, H. J. Silverstein, C. R. Wiebe, J. G. Cheng, and H. D. Zhou, Phys. Rev. B {\bf 92}, 140407(R) (2015).

\bibitem{Er2Sn2O7-1}
M. Shirai, R. S. Freitas, J. Lago, S. T. Bramwell, C. Ritter, and I. \v{Z}ivkovi\'{c}, Phys. Rev. B {\bf 96}, 180411(R) (2017).

\bibitem{Er2Sn2O7-2}
S. Petit, E. Lhotel, F. Damay, P. Boutrouille, A. Forget, and D. Colson, Phys. Rev. Lett. {\bf 119}, 187202 (2017).

\bibitem{Er2Pt2O7-1}
Y. Q. Cai, Q. Cui, X. Li, Z. L. Dun, J. Ma, C. dela Cruz, Y. Y. Jiao, J. Liao, P. J. Sun, Y. Q. Li, J. S. Zhou, J. B. Goodenough, H. D. Zhou, and J. G. Cheng, Phys. Rev. B {\bf 93}, 014443 (2016).

\bibitem{Er2Pt2O7-2}
A. M. Hallas, J. Gaudet, N. P. Butch, G. Xu, M. Tachibana, C. R. Wiebe, G. M. Luke, and B. D. Gaulin, Phys. Rev. Lett. {\bf 119}, 187201 (2017).

\bibitem{AC}
Z. L. Dun, M. Lee, E. S. Choi, A. M. Hallas, C. R. Wiebe, J. S. Gardner, E. Arrighi, R. S. Freitas, A. M. Arevalo-Lopez, J. P. Attfield, H. D. Zhou, and J. G. Cheng, Phys. Rev. B {\bf 89}, 064401 (2014).

\bibitem{kappa-1}
F. B. Zhang, Q. J. Li, Z. Y. Zhao, C. Fan, S. J. Li, X. G. Liu, X. Zhao, and X. F. Sun, Phys. Rev. B {\bf 89}, 094403 (2014).

\bibitem{kappa-2}
S. J. Li, Z. Y. Zhao, C. Fan, B. Tong, F. B. Zhang, J. Shi, J. C. Wu, X. G. Liu, H. D. Zhou,
X. Zhao, and X. F. Sun, Phys. Rev. B {\bf 92}, 094408 (2015).

\bibitem{kappa-3}
C. C. Gu, Z. Y. Zhao, X. L. Chen, M. Lee, E. S. Choi, Y. Y. Han, L. S. Ling, L. Pi, Y. H. Zhang, G. Chen, Z. R. Yang, H. D. Zhou, and X. F. Sun, Phys. Rev. Lett. {\bf 120}, 147204 (2018).

\bibitem{IR}
R. D. Shannon, Acta. Cryst. {\bf A32}, 751 (1976).

\bibitem{Er2Ti2O7-XRD}
Y. Tabira, R. L. Withers, L. Minervini, and R. W. Grimes, J. Solid State Chem. {\bf 153}, 16 (2000).

\bibitem{Er2Ti2O7-10}
P. Bonville, S. Petit, I. Mirebeau, J. Robert, E. Lhotel, and C. Paulsen, J. Phys.: Condens. Matt. {\bf 25}, 275601 (2013).

\bibitem{Er2Ti2O7-crystal growth}
Q. J. Li, L. M. Xu, C. Fan, F. B. Zhang, Y. Y. Lv, B. Ni, Z. Y. Zhao, X. F. Sun, J. Cryst. Growth {\bf 377}, 96 (2013).

\bibitem{Er2Ti2O7-9}
P. Dasgupta, Y. Jana, D. Ghosh, Solid State Commun. {\bf 139}, 424 (2006).

\bibitem{Er2Sn2O7-3}
K. Matsuhira, Y. Hinatsu, K. Tenya, H. Amitsuka, and T. Sakakibara, J. Phys. Soc. Jpn. {\bf 71}, 1576 (2002).

\bibitem{Er2Sn2O7-4}
P. M. Sarte, H. J. Silverstein, B. T. K. Van Wyk, J. S. Gardner, Y. Qiu, H. D. Zhou, and C. R. Wiebe, J. Phys.: Condens. Matt. {\bf 23}, 382201 (2011).

\bibitem{Mydosh-1}
J. A. Mydosh, \textit{Spin Glass: An Experimental Introduction} (Taylor and Francis, London, 1993).

\bibitem{Mydosh-2}
J. A. Mydosh, Rep. Prog. Phys. {\bf 78}, 052501 (2015).

\bibitem{PrRhSn3}
V. K. Anand, D. T. Adroja, and A. D. Hillier, Phys. Rev. B {\bf 85}, 014418 (2012).

\bibitem{CeNiCu}
N. Marcano, J. C. G\'{o}mez Sal, J. I. Espeso, L. Fern\'{a}ndez Barqu\'{\i}n, and C. Paulsen, Phys. Rev. B {\bf 76}, 224419 (2007).

\bibitem{zv-1}
J. Souletie and J. L. Tholence, Phys. Rev. B {\bf 32}, 516 (1985).

\bibitem{zv-2}
K. H. Fischer and J. Hertz, {\it Spin Glasses} (Cambridge University Press, 1991).

\bibitem{Er2GaSbO7}
H. W. J. Bl\"{o}te, R. F. Wielinga, and W. J. Huiskamp, Physica {\bf 43}, 549 (1969).

\bibitem{Berman}
R. Berman, {\it Thermal Conduction in Solids} (Oxford University Press, Oxford, 1976).

\bibitem{specular-reflection}
M. Sutherland, D. G. Hawthorn, R. W. Hill, F. Ronning, S. Wakimoto, H. Zhang, C. Proust, E. Boaknin, C. Lupien, L. Taillefer, R. Liang, D. A. Bonn, W. N. Hardy, Robert Gagnon, N. E. Hussey, T. Kimura, M. Nohara, and H. Takagi, Phys. Rev. B {\bf 67}, 174520 (2003).

\bibitem{kappa-4}
X. F. Sun, A. A. Taskin, X. Zhao, A. N. Lavrov, and Y. Ando, Phys. Rev. B {\bf 77}, 054436 (2008).

\bibitem{kappa-5}
Q. J. Li, Z. Y. Zhao, H. D. Zhou, W. P. Ke, X. M. Wang, C. Fan, X. G. Liu, L. M. Chen, X. Zhao, and X. F. Sun, Phys. Rev. B {\bf 85}, 174438 (2012).

\bibitem{kappa-6}
Q. J. Li, Z. Y. Zhao, C. Fan, F. B. Zhang, H. D. Zhou, X. Zhao, and X. F. Sun, Phys. Rev. B {\bf 87}, 214408 (2013).

\bibitem{Er2-xYxTi2O7-theory-1}
V. S. Maryasin and M. E. Zhitomirsky, Phys. Rev. B {\bf 90}, 094412 (2014).

\bibitem{Er2-xYxTi2O7-theory-2}
A. Andreanov and P. A. McClarty, Phys. Rev. B {\bf 91}, 064401 (2015).

\bibitem{Er2-xYxTi2O7-experiment-2}
J. Gaudet, A. M. Hallas, D. D. Maharaj, C. R. C. Buhariwalla, E. Kermarrec, N. P. Butch, T. J. S. Munsie, H. A. D\c{a}bkowska, G. M. Luke, and B. D. Gaulin, Phys. Rev. B {\bf 94}, 060407(R) (2016).

\bibitem{CGS}
E. C. Andrade, J. A. Hoyos, S. Rachel, and M. Vojta, Phys. Rev. Lett. {\bf 120}, 097204 (2018).

\bibitem{NaCaCo2F7-1}
J. W. Krizan and R. J. Cava, Phys. Rev. B {\bf 89}, 214401 (2014).

\bibitem{NaSrCo2F7-1}
J. W. Krizan and R. J. Cava, J. Phys.: Condens. Matt. {\bf 27}, 296002 (2015).

\bibitem{NaCaCo2F7-2}
K. A. Ross, J. W. Krizan, J. A. Rodriguez-Rivera, R. J. Cava, and C. L. Broholm, Phys. Rev. B {\bf 93}, 014433 (2016).

\bibitem{NaCaCo2F7-3}
K. A. Ross, J. M. Brown, R. J. Cava, J. W. Krizan, S. E. Nagler, J. A. Rodriguez-Rivera, and M. B. Stone, Phys. Rev. B {\bf 95}, 144414 (2017).

\end{thebibliography}
\end{document}